\documentclass[%
onecolumn,
superscriptaddress,
amsmath,
amssymb,
aps, 
pra,
floatfix,
11pt,
]{revtex4-2}

\usepackage{graphicx}
\usepackage{dcolumn}
\usepackage{bm}

\input{preamble.tex}

\begin{document}

\hypersetup{
  linkcolor=darkrust,
  citecolor=seagreen,
  urlcolor=darkrust,
  pdfauthor=author,
}


\title{Competing Mechanisms at Vibrated Interfaces of Density-Contrast Fluids} 

\author{Tianyi Chu}
\email{Contact author: tchu72@gatech.edu}
\affiliation{School of Computational Science \& Engineering, Georgia Institute of Technology, Atlanta, Georgia 30332-0365, USA}

\author{Benjamin Wilfong}
\affiliation{School of Computational Science \& Engineering, Georgia Institute of Technology, Atlanta, Georgia 30332-0365, USA}

\author{Timothy Koehler}
\affiliation{Engineering Sciences Center, Sandia National Laboratories, P.O.\ Box 5800, Albuquerque, New Mexico 87185-0840, USA}

\author{Ryan M.\ McMullen}
\affiliation{Engineering Sciences Center, Sandia National Laboratories, P.O.\ Box 5800, Albuquerque, New Mexico 87185-0840, USA}

\author{Spencer H.\ Bryngelson}
\affiliation{School of Computational Science \& Engineering, Georgia Institute of Technology, Atlanta, Georgia 30332-0365, USA}

\begin{abstract}
Fluid--fluid interfacial instability and subsequent fluid mixing are ubiquitous in nature and engineering.
The hydrodynamic instability of fluid interfaces has long centered on the pressure gradient-driven long-wavelength Rayleigh--Taylor instability and the resonance-induced short-wavelength Faraday instability.
However, neither instability alone can explain the dynamics when both mechanisms are present.
We identify a previously unseen multi-modal instability emerging from their coexistence.
When the denser fluid is polydimethylsiloxane, the mixed region at a high density contrast (Atwood number=0.9) spans a vibration amplitude range approximately twice the gravitational acceleration.
Using Floquet stability analysis, we show how vibrations govern transitions between the RT and Faraday instabilities, leading to contention between these instabilities rather than resonant enhancement.
The initial transient growth is represented by the exponential modal growth of the most unstable Floquet exponent, along with its accompanying periodic behavior.
Direct numerical simulations validate these findings and track interface breakup into the multiscale and nonlinear regimes.
Specifically, we show that growing RT modes nonlinearly suppress Faraday responses even when the initial growth rate of the Faraday instability is 3.63 times that of RT, so a bidirectional competition hinders their sustained coexistence.
\end{abstract}

\maketitle


\section{Introduction} 


The interface separating phases in multiphase fluid systems is often subjected to deformation due to internal density differences or external vibrations.
These deformations manifest in various forms, including
liquid dripping~\citep{wilson1988slow,clanet1999transition,montanero2020dripping},
bubble injection~\citep{Bleich1956,chen1997instability,yang2007bubble}, 
tip streaming~\citep{eggleton2001tip,collins2008electrohydrodynamic,montanero2020dripping},
and surface waves~\citep{miles1990parametrically}.
Such phenomena are ubiquitous, occurring in natural processes such as mantle plumes and cryospheric vibrations, as well as in engineering applications like atomization and inertial confinement fusion~\citep{craxton2015direct,betti2016inertial}.
The unstable growth of these interface deformations eventually leads to interface breakup and fluid mixing.
Linear stability analysis predicts two primary hydrodynamic instability mechanisms: the pressure gradient-driven fluid mixing mechanism, triggering the Rayleigh--Taylor (RT) instability~\citep{rayleigh1882investigation,taylor1950instability},
and the resonance mechanism, triggering a Faraday instability~\citep{faraday1831xvii}.
Over a century of research has focused on studying these fundamental hydrodynamic instabilities separately, as each is central to a vast range of applications.


The analysis of RT instability originated with investigations into the growth of small-amplitude interfacial waves between two inviscid fluids of differing densities~\citep{rayleigh1882investigation,taylor1950instability}.
Viscous effects in the stable configuration were first considered by \citet{harrison1908influence}, 
and later, \citet{bellman1954effects} examined the combined influence of viscosity and surface tension on the unstable regime.
A comprehensive and widely acclaimed treatise on RT instability can be found in~\citet{chandrasekhar1961hydrodynamic}.
Building on this foundation, \citet{plesset1974viscous} introduced simplified physical models applicable in the asymptotic limits of large and small contrasts in density and viscosity.
For detailed reviews of RT instability theory and its developments, readers are referred to \citet{sharp1983overview} and \citet{kull1991theory}.
More recently, \citet{piriz2006} derived a simple yet highly accurate analytical expression for RT instability in nonideal fluids.

When vibration amplitude exceeds a critical threshold, the flat interface destabilizes through resonant Faraday instability, generating standing surface waves.
This phenomenon has garnered significant attention and extensive study; see an early review by \citet{miles1990parametrically}.
\citet{benjamin1954stability} showed that the stability of the free surface of an ideal fluid can be analyzed using a system of Mathieu equations.
Although widely adopted, this idealized model is not sufficiently realistic to yield results that match experimental observations~\citep{fauve1992parametric}.
This discrepancy shows the role of viscosity in interfacial stability.
Using Floquet analysis, \citet{kumar1994parametric} showed that even when linear damping is included, the Mathieu equation fails to represent the dynamics of viscous fluids.
They demonstrated that viscosous effects modify the wavelength selection and distort the tongue-shaped stability regions.
\citet{kumar1996} extended this analysis to shallow fluids with free surfaces, predicting a series of bicritical points for subharmonic and harmonic responses.
Despite its limitations in representing viscous effects, \citet{wright2000numerical} numerically demonstrated that the modified Mathieu equation accurately predicts the temporal growth of Faraday oscillations in inviscid fluids.
Beyond the waveforms observed in the above two-dimensional analyses and experiments, \citet{panda2025marangoni} recently demonstrated numerically that Faraday instability can give rise to more complex patterns when the domain is extended to three dimensions.

Despite their well-established nature and pervasive observations, a comprehensive understanding of the interface behavior when the long-wavelength RT and short-wavelength Faraday instabilities coexist is lacking.
This knowledge gap is important because vibrations and density contrast occur simultaneously in many realistic situations, including vibrated multiphase systems, industrial mixing processes, and fluids in spacecraft propellant tanks.
For example, vibrated levitating liquids represent a compelling case in which these instabilities interact in a nontrivial way~\citep{apffel2020floating}.
Faraday waves inherently link to RT instability when the oscillation amplitude exceeds gravitational acceleration~\citep{kumar2000mechanism,wright2000numerical}.
In such regimes, the interface undergoes the Faraday instability throughout each oscillation cycle, while experiencing RT mechanisms during the upward acceleration phase.
Notably, \citet{dinesh2023pattern} experimentally demonstrated the emergence of an electrostatic resonance resembling an RT-like instability, coexisting with the standard Faraday instability.
A fundamental understanding of these coupled mechanisms is essential for developing effective strategies to control interfacial dynamics in such complex systems.
\citet{wolf1969dynamic,wolf1970dynamic} showed via experiment that unstable viscous RT waves can be stabilized  \textit{without} generating standing Faraday waves by oscillating the container at specific amplitudes and frequencies.
Later studies supported this conclusion through theoretical arguments, and a partial theory was developed~\citep{Troyon1971, lapuerta2001control, piriz2010dynamic, sterman2017rayleigh, pototsky2016faraday}.
However, the coexistence of unstable RT and Faraday modes, which can meaningfully influence the interface dynamics, has been largely overlooked.
The complex interactions and physical mechanisms emerging from this coexistence remain mysterious.
This work addresses this gap and provides a comprehensive view of interfacial dynamics under combined density contrast and vibrations, while also accounting for viscosity and surface tension effects.



\begin{figure}[ht]
    \centering
    \includegraphics[scale=1]{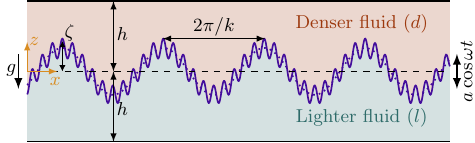}
    \caption{
        The two-fluid interface with long-wavelength RT and short-wavelength Faraday instabilities. 
    }
    \label{f:overview}
\end{figure}

We focus on two-dimensional analyses and simulations to gain clearer insight into the underlying mechanisms.
\Cref{f:overview} shows a schematic of the two-fluid interface, where both long-wavelength RT and short-wavelength Faraday waves \textit{coexist}. 
In \cref{background}, we perform a linear Floquet stability analysis to characterize the temporal evolution of the interface in the linear regime.
Assuming horizontal homogeneity, we decompose the interface into individual wave numbers, enabling separate investigation of both long-wavelength RT and short-wavelength Faraday modes.
\Cref{numerical method} describes the numerical approach for two-dimensional, scale-resolved simulations of the two-phase fluid system.
These simulations validate the linear theory and represent the system evolution into multiscale and nonlinear regimes, as presented in \cref{results}.
We also explore the oscillation--frequency-amplitude phase space to characterize the transition of the dominant instability.
\Cref{conclusion} discusses the limitations of the study and summarizes the findings on the coexistence and competition between RT and Faraday instabilities.



\section{Theoretical Background}\label{background}

\subsection{Governing equations}
Consider the interface between two immiscible and incompressible fluids: a denser and lighter fluid, denoted by subscripts $\left(\cdot\right)_d$ and $\left(\cdot\right)_l$. 
The fluids are confined between two infinite horizontal plates separated by a distance of $2h$ and subject to constant gravitational acceleration and an oscillatory vertical acceleration
\begin{equation}
    \tilde{g}(t) = g^*g + a \cos (\omega t+\varphi_0),
\end{equation}
where $a$ is the oscillatory amplitude, $\omega$ is the corresponding frequency, and $\varphi_0$ is the initial phase.
Positive gravity, $g^*=1$, applies when the denser fluid is at the domain bottom, and negative gravity, $g^*=-1$, applies when the denser fluid is on top.
\Cref{f:overview} shows the two-fluid interface subjected to a harmonic vibration.
Within each fluid layer $j$, the motion of the fluid is governed by the incompressible Navier--Stokes equations,
    \begin{align}\label{eqn:NS}
    \rho_j \left[\partial_t +\left(\vb*{u}_j\cdot \vb*{\nabla}\right) \right]\vb*{u}_j=-\grad p_j-\rho_j \tilde{g}\vb*{e}_z +\mu_j \grad^2\vb*{u}_j,\quad 
    \nabla\cdot \vb*{u}_j =0.
\end{align}
To systematically analyze the interfacial dynamics, we nondimensionalize the governing equations using $h$ as the characteristic length and $g$ as the reference gravitational acceleration.
Density, $\rho$, viscosity, $\mu$, and surface tension, $\sigma$ are scaled using appropriate reference values.
The resulting dimensionless variables, denoted by $(\cdot)^*$, are summarized in \cref{dimensionless_quantity}.

\begin{table}[ht!]
    \centering
    \caption{Dimensionless quantities.}\label{dimensionless_quantity}
    {\setlength{\tabcolsep}{3pt}\small
    \begin{tabular}{l l}
     Dimensionless quantity  & Quantity name       \\ \midrule
        $t_c=\sqrt{h/g}$    & Gravitational time  \\
      $\At=(\rho_d-\rho_l)/(\rho_d+\rho_l)$ & Atwood number \\
     $\eta = \mu_l/\mu_d$ & Viscosity ratio \\
     $C= \nu_d/\sqrt{gh^3}$ & Viscous-to-gravitational\\
     & force ratio\\
     $\mathrm{Bd} = \rho_d gh^2/\sigma_d$ & Bond number\\
       $a^*=a/g$            & Acceleration  \\
     $\zeta^*=\zeta/h$             & Interface displacement   \\
      $k^*=kh$             & Wavenumber   \\
        $t^*=t/t_c$              & Time\\
       $\gamma^*=\gamma t_c$            & Growth rate  \\
        $\omega^*=\omega t_c$ & Oscillatory frequency  
    \end{tabular}
    }
\end{table}
Eq.~\cref{eqn:NS} can therefore be written as
    \begin{align}\label{eqn:full NS}
    \left[\partial_{t^*} +\left(\vb*{u}^*_j\cdot \vb*{\nabla}^*\right)\right] \vb*{u}^*_j&=-\grad^* p^*_j- \tilde{g}^*\vb*{e}_z +C_j {\boldsymbol{\Delta}^*}\vb*{u}^*_j,\quad 
    \grad^*\cdot \vb*{u}_j^* =0,
\end{align}
where $C_{d}=C$, and $C_l = (1-\At)/(\eta(1+\At))$.
We can decompose the flow state around the equilibrium-state solution of the Navier--Stokes equations \cref{eqn:full NS} as
\begin{align}
         \vb*{u}^*_j=\vb*{U}^*_j+ {\vb*{u}^*_j}',\quad  p^*_j={P}^*_j+  {p^*_j}',
\end{align}
where $(\vb*{U}^*_j,\, P^*_j)=(0,- \tilde{g}^*\vb*{e}_z )$
and $(\cdot)'$ denotes the small fluctuating components.
The linearized governing equations for the fluctuations are
\begin{align}\label{eqn:NS fluctuating}
     \partial_{t^*} {\vb*{u}_j^*}'=-\grad^* {p^*_j}' +C_j {\boldsymbol{\Delta}^*}{\vb*{u}_j^*}', \quad 
    \grad^* \cdot {\vb*{u}_j^*}' =0.
\end{align}
Applying the operator $\vb*{e}_z\cdot \grad^*\times\grad^*\times$ to Eq.~\cref{eqn:NS fluctuating} eliminates the horizontal velocity components, yielding
\begin{align}\label{eqn:NS fluctuating_w}
    \left(\partial_{t^*}-C_j{\boldsymbol{\Delta}^*}\right){\boldsymbol{\Delta}^*} {w_j^*}'=0,
\end{align}
At the interface $z^*=\zeta^*$, continuity of the vertical velocity perturbation, ${w_j^*}'$, and the tangential stress yields
\begin{align}\label{eqn:interface_cont}
    \delta \left[{w_j^*}',\, \partial_{z^*}\left( {w_j^*}'\right),\, C_j\left(\boldsymbol{\Delta}^*-2\partial_{z^*z^*}\right){w_j^*}'\right]=0,
\end{align}
where $\delta$ represents the jump across the interface. 
The linearized kinematic boundary condition gives
\begin{align}\label{eqn:kinematic}
    \partial_{t^*} {\zeta}^* &={w}^*\vert_{z^*=0}.
\end{align}
The jump condition for pressure across the interface,
\begin{align}\label{eqn:pressure jump}
\delta p^* &= 2(C_l-C_d) \left(\partial_{z^*}{w}^*\right)\vert_{z^*=0}-\frac{2\mathrm{At}}{1+\mathrm{At}}\tilde{g}^*\zeta^*+\frac{1}{\mathrm{Bd}}\left(\boldsymbol{\Delta}^*-\partial_{z^*z^*}\right)\zeta^*,
\end{align}
explicitly represents the effect of vibration.
Together, Eqs.~\Crefrange{eqn:NS fluctuating_w}{eqn:pressure jump} govern the linearized interfacial dynamics. 
We further assume that the interface profile, $\zeta^*$, remains well-defined and single-valued during the vibration.
Due to the horizontal homogeneity, we use a Fourier transform in the horizontal direction,
\begin{align}
    \left[{w^*_j}'({x}^*,z^*, t^*),\, \zeta^*({x}^*, t^*)\right] = \frac{1}{2\pi}\int_{-\infty}^{\infty}
     \left[\hat{w}^*_{j}(k^*,z^*,t^*),\, \hat{\zeta}^*(k^*,t^*)\right]\rme^{\rmi k^* {x}^*}\,\dd k^*. \label{eqn:decomposition_w} 
\end{align}
to expand the velocity and interface displacement in the above governing equations.
Eq.~\cref{eqn:decomposition_w} decomposes waveforms into different length scales, enabling separate analysis of RT and Faraday instabilities.
The dynamics introduced by the parametric source are subsequently analyzed using a Floquet analysis, as described next.

\subsection{Linear Floquet stability analysis}

The time-periodic variation in gravity is suspected to be the driving source of instability, thereby motivating a Floquet analysis~\citep{floquet1883equations}.
Consequently, the temporal evolution of the vertical interface displacement, $\zeta^*(t^*)$, is expected to exhibit both exponential growth or decay, modeled by $\rme^{\gamma^* t^*}$, and superimposed oscillatory fluctuations, captured by the periodic term $\sum_{n} \hat{\zeta}^*_n \rme^{\rmi n \omega^* t^*}$.
At each horizontal wavenumber, this response can be decomposed into modal and periodic parts, yielding
\begin{align}\label{eqn:Floequet_form}
    & \left[\hat{w}^*_{j}(k^*,z^*,t^*),\, \hat{\zeta}^*(k^*,t^*)\right]= 
    \underbrace{\rme^{\gamma^* t^*}}_{\text{Modal}}\,
    \underbrace{\sum_{n=-\infty}^{\infty}\left[\hat{w}_{j,n}^*(k^*,z^*),\,\hat{\zeta}_n^*(k^*)\right]\rme^{\rmi n \omega^* (t^*+t_0^*)}}_{\text{Periodic components}},
\end{align}
where $\gamma^*$ is the Floquet exponent, $t_0^*$ is the time shift associated with the initial phase $\varphi_0$, and $n$ is the intger index of the harmonics \citep{kumar1994parametric,kumar1996}.
A positive real Floquet exponent indicates instability, driving the system toward interface breakup.
Substituting Eqs.~\cref{eqn:decomposition_w,eqn:Floequet_form} into Eq.~\cref{eqn:NS fluctuating_w} gives 
\begin{align}\label{eqn:governing_w}
    &\left[\gamma_n^*-C_jD_{zz}^{(k)}\right]D_{zz}^{(k)}\hat{w}^*_{n,j}=0,
\end{align}
where $\gamma_n^*\equiv \gamma^* +\rmi n\omega^*$ is the total Floquet exponent for the $n$th harmonics, and 
$D_{zz}^{(k)} \equiv \partial_{z^*z^*}-{k^*}^2$.
The general solution of this dispersion relation is
\begin{align}\label{eqn:w_genearal}
     \hat{w}^*_{n,j} &=  a_{n,j} \rme^{k^* z^*}+b_{n,j} \rme^{-k^* z^*} +c_{n,j} \rme^{ k^*q_{n,j} z^*}+d_{n,j} \rme^{- k^*q_{n,j} z^*},
\end{align}
where
$q_{n,j}\equiv\sqrt{1+\gamma_n^*/(C_jk^*)}$.
No-slip boundary conditions at the two plates yield
\begin{align}\label{eqn:BC_vertical}
    \hat{w}^*_{n,d} = \partial_{z^*} \hat{w}^*_{n,d} = 0 \quad &\text{at} \quad z^*=-1, \\
    \hat{w}^*_{n,l} = \partial_{z^*} \hat{w}^*_{n,l} = 0 \quad &\text{at} \quad z^*=1.
\end{align}
The jump conditions at the interface in Eqs.~\cref{eqn:interface_cont,eqn:kinematic} can be written as 
\begin{align}
     \left[1,\,\partial_{z^*},\, \eta \left(\partial_{z^*z^*}+{k^*}^2\right)\right]\hat{w}^*_{n,l}&= \left[1,\,\partial_{z^*},\,  \left(\partial_{z^*z^*}+{k^*}^2\right)\right]\hat{w}^*_{n,d},\\
    \gamma_n^* \hat{\zeta}^*_n &=\hat{w}^*_n\vert_{z^*=0}.\label{eqn:BC8}
\end{align}
Substituting the general solution in Eq.~\cref{eqn:w_genearal} into the above 8-equation boundary condition, we find
\begin{align}
     &{\vb*{Q}_n}
\mqty(a_{n,d}, b_{n,d},c_{n,d},d_{n,d},a_{n,l}, b_{n,l},c_{n,l},d_{n,l})^\transpose  = 
     \mqty(\gamma^*_n \hat{\zeta}^*_n, 0,0,0,0,0,0,0)^\transpose.
\end{align}
The full expression of the matrix $\vb*{Q}_n$ is given in Eq.~\cref{eqn:Q} in the appendix.
When the depth $h \to \infty$, the vanishing velocity gives $b_{n,d}=d_{n,d}=a_{n,l}=c_{n,l}=0$.

The jump condition in Eq.~\cref{eqn:pressure jump} directly relates the displacements at different orders of harmonics, $\hat{\zeta}_n^*$. 
The resulting linear system can be expressed as
\begin{align}
 &\qquad \qquad \qquad \left[ \underbrace{ \mqty(\ddots & \vdots &  \vdots & \vdots & \vdots \\
    \cdots  & A_{-1}& 0&0  & \cdots\\
    \cdots  & 0& A_{0}&0  & \cdots\\
    \cdots  & 0& 0&A_{1}  & \cdots\\
    \vdots & \vdots & \vdots &  \vdots &\ddots
    )}_{\vb*{A}}
- a^* \underbrace{ \mqty(\ddots & \vdots & \vdots  & \vdots &\vdots \\
    \cdots  & 0& 1&0  & \cdots\\
    \cdots  & 1& 0&1  & \cdots\\
    \cdots  & 0& 1&0  & \cdots\\
    \vdots & \vdots & \vdots & \vdots  &\ddots
    )}_{\vb*{B}} \right]
       \underbrace{ \mqty(\vdots\\ \hat{\zeta}_{-1}^*\\\hat{\zeta}_0^*\\\hat{\zeta}_1^*\\\vdots)}_{\hat{\vb*{\zeta}}^*} =0, \quad \text{where} \label{eqn:General EVD}\\
        &A_n(\gamma^*;k^*) \equiv 
        2k^*\left(a_{n,d}-b_{n,d}+c_{n,d}q^*_{n,d}-d_{n,d}q^*_{n,d}\right)\left(\gamma^*_n/(C{k^*}^2)+3C(1-\eta)(1+\At)/(2\At) \right) \label{eqn:An} \\ 
        &-2C(1-\eta)k^*(1+\At)/(2\At)\left(a_{n,d}-b_{n,d}+c_{n,d}{q}_{n,d}^{*^3}-d_{n,d}{q}_{n,d}^{*^3}\right)
    + 2(g^*+{k^*}^2(1+\At)/(2\mathrm{At}\mathrm{Bd})). \nonumber
\end{align}
Eqs.~\cref{eqn:General EVD,eqn:An} provide a comprehensive analysis of interfacial instabilities, accounting for viscosity effects, surface tension, internal density differences, and external vibrations.
For each wavenumber, the solution of the optimization problem
\begin{align}
    \gamma^*_{U}= \argmax_{\det{\vb*{A}(\gamma^*;\, k^*)-a^*\vb*{B}} = 0}  \mathrm{Re}\{\gamma^*\}.
    \label{eqn:Det}
\end{align}
yields the corresponding most unstable Floquet exponent.
The periodic components at different harmonics, $\hat{\vb*{\zeta}}^*$, can be then found as the null space of $(\vb*{A}-a^*\vb*{B})$.
Equation~\cref{eqn:Det} shows a generalized analysis of the hydrodynamic instability in fluid mixing, covering a range of configuration parameters, including viscosity, surface tension, vibration, and density differences, exhibiting rich mechanical differences.
In the static limit ($a^*\to 0$), Eq.~\cref{eqn:Det} becomes
\begin{align}
 \gamma^*_{U}= \argmax_{A_0(\gamma^*;\,k^*)= 0}  \mathrm{Re}\{\gamma^*\},
\end{align}
as only $n=0$ survives the simplification of Eq.~\cref{eqn:Floequet_form}. 
This form recovers the classical RT dispersion relation~\citep{chandrasekhar1961hydrodynamic}.
The critical acceleration, $a^*_c$, for neutral stability is obtained by constraining the displacement in Eq.~\cref{eqn:Det} to exhibit pure sinusoidal harmonic~(H), $\gamma^*_{U} = 0$, or subharmonic~(S), $\gamma^*_{U} = \rmi \omega^*/2$, responses,
which recovers the generalized eigenvalue problem,
\begin{align}\label{eqn:EVD}
    \vb*{A}(\gamma^*_{U}=0\,\,\text{or}\,\, \rmi \omega^*/2;\, k^*) \hat{\vb*{\zeta}}^*=a_c^*\vb*{B} \hat{\vb*{\zeta}}^*,
\end{align}
for the Faraday instability~\citep{kumar1994parametric,kumar1996}.
Beyond identifying the stability boundary, Eq.~\cref{eqn:Floequet_form} enables the prediction of transient wave dynamics at a given wavenumber $k^*$ within the unstable regime. 
In practice, Eq.~\cref{eqn:General EVD} is truncated at $n=10$, resulting in a $22\times22$ linear problem.
Including higher-order harmonics does not visibly affect the results but may introduce numerical instabilities.
Together, these analyses enable the dissection of the mechanics as the system transitions toward nonlinear interface breakup.
While standard Floquet exponents, often obtained through the eigendecomposition of the monodromy matrix, are not unique, they collectively describe nonmodal growth.
Using numerical simulations, we demonstrate that the modal growth of the most unstable Floquet exponent, $\gamma^*_{U}$, combined with the periodic components, can accurately predict the initial transient dynamics.



\section{Numerical method}\label{numerical method}

We perform high-fidelity 2D simulations using MFC, a GPU-accelerated compressible solver for multi-component, multiphase flows~\citep{bryngelson19_cpc}.
MFC is selected for its computational efficiency, state-of-the-art GPU acceleration~\citep{radhakrishnan24,wilfong252}, and ability to handle numerically challenging problems, including multi-component high-density ratio flows with interfacial tension effects that we focus on here.
For example, the present method is validated against analytical solutions for Rayleigh--Taylor and Faraday instabilities, showing deviations of less than 1$\%$ from exact (see \cref{results}).


Here, we use the six-equation diffuse interface model~\citep{Saurel2008} to describe the multiphase system, which we augmented with a surface tension formulation~\citep{schmidmayer2017}.
The governing equations have the form
\begin{equation}
  \pdv{\vb*{q}}{t}+\grad\cdot\vb*{F}(\vb*{q})+\vb*{h}(\vb*{q})\grad\cdot \vb*{u} = \vb*{s}(\vb*{q}),
  \label{eq:mfc_cons}
\end{equation}
for which
\begin{align}
    \vb*{q} &= \begin{bmatrix}
        \alpha_1 \\
        \alpha_1 \rho_1 \\
        \alpha_2 \rho_2 \\
        \rho \vb*{u} \\
        \alpha_1\rho_1e_1 \\
         \alpha_2\rho_2e_2\\
         \rho E+\epsilon_0
    \end{bmatrix}, 
    \vb*{F} = \begin{bmatrix}
        \alpha_1 \vb*{u} \\
        \alpha_1 \rho_1 \vb*{u} \\
        \alpha_2 \rho_2 \vb*{u} \\
        \rho \vb*{u} \vb*{u}   + p\vb*{I} - \vb*{T}+\vb*{\Omega}\\
        \alpha_1\rho_1e_1\vb*{u}\\
        \alpha_2\rho_1e_2\vb*{u}\\
        (\rho E + p+\epsilon_0) \vb*{u} - \vb*{T} \cdot \vb*{u}+\vb*{\Omega}\cdot\vb*{u} \\
    \end{bmatrix}, 
    \vb*{h} = \begin{bmatrix}
        -\alpha_1 \\
       0 \\
       0 \\
       \vb*{0} \\
     \alpha_1p_1\\
     \alpha_2 p_2\\
     0
    \end{bmatrix}, 
    \vb*{s} = \begin{bmatrix}
       \mu\delta p \\
       0 \\
        0 \\
       -\rho \tilde\bg \\
        -\mu p_I\delta p-\alpha_1\vb*{T}_1:\grad \vb*{u}_1\\
        \phantom{-}\mu p_I\delta p-\alpha_2\vb*{T}_2:\grad \vb*{u}_1\\
        -\rho(\bu\cdot\tilde\bg)
    \end{bmatrix}.
    \label{eq:mfc_cons2}
\end{align}
For each fluid $j$, $\alpha_j$ is the volume fraction, $e_j$ is the internal energy, and
\begin{gather}
    \vb*{T}_j \equiv \mu_j\left[\grad \vb*{u}+(\grad \vb*{u})^\transpose -\frac{2}{3}\left(\grad\vdot \vb*{u}\right)\vb*{I}\right]
    \label{eq:viscous_stress}
\end{gather} 
is the viscous stress tensor.
The equations are closed by the usual set of mixture rules
\begin{align}
    \sum_{j}\alpha_j = 1, \quad 
    \rho = \sum_j\rho_j\alpha_j, \quad
    \mu = \sum_j\mu_j\alpha_j, \quad 
    \text{and} \quad
    \vb*{T} = \sum_j \vb*{T}_j \alpha_j.
\end{align}
The total mixture energy $E$ is
\begin{gather}
    E = \sum_{j}Y_je_j + ||\bu||^2/2,
\end{gather}
where $Y_j = \alpha_j\rho_j/\rho$ are the mass fractions of each phase.
Surface tension and body forces are applied only to the total mixture energy.
The capillary stress tensor $\bOmega$ is given by
\begin{equation}
    \bOmega = -\sigma\left(\lVert\nabla \alpha_1\rVert\bI - \frac{\nabla \alpha_1 \otimes \nabla \alpha_1}{\lVert\nabla \alpha_1\rVert}\right),
\end{equation}
 and $\epsilon_0 = \sigma \lVert \nabla \alpha \rVert$ is the capillary energy.
The system is solved via a high-order accurate finite-volume method and Runge--Kutta temporal discretization.
The numerical method closely follows that of \citet{Coralic2014} and is described in detail in \citet{bryngelson19_cpc} and \citet{wilfong252}.
Closure is provided by the stiffened gas equation of state
\begin{equation}
    e_j = \frac{p_j + \gamma_j \pi_{\infty,j}}{\left( \gamma_j - 1\right) \rho_j}.
\end{equation}
The fluid parameters for ratio of specific heats, $\gamma_j$,  and liquid stiffness, $\pi_{\infty, j}$, enable faithful modeling of liquids and gasses~\cite{menikoff1989}.

{\normalfont\bfseries\centering Numerical simulations \, }
The 2D computational domain spans four wavelengths of interest in width, with each fluid layer having a depth of $h = \SI{0.013}{m}$.
A uniform Cartesian mesh resolves the shortest wavelength with 192 grid cells, which is sufficient to accurately capture the interface between the two fluids. This resolution has been validated through a grid convergence study conducted with up to 288 grid cells.
Periodic and no-slip boundary conditions are applied at the horizontal and vertical boundaries, respectively.
The interface is initially perturbed by a height that is $1\%$ of the wavelength, and hydrodynamic equilibrium is achieved at the interface with a pressure of $\SI{100}{\kilo\pascal}$.
For demonstration purposes, we consider polydimethylsiloxane (PDMS) as the more dense fluid, which has been used in experimental studies of vibrated fluids~\citep{o2012bubble}.
PDMS is characterized as a Newtonian fluid with density $\rho_d = \SI{950}{\kilo\gram\per\meter\cubed}$, kinematic viscosity of $\nu_d = \SI{2e{-5}}{\meter\squared\per\second}$, and surface tension coefficient of $\sigma_d = \SI{2.06e{-2}}{\newton\per\meter}$.
The stiffened gas equation of state parameters $(\gamma,\pi_\infty)=(3.49,2.78\times10^8)$ for PDMS are selected to match the speed of sound $\SI{1011}{\metre/\second}$~\citep{zhang2021} and the internal energy of water at standard temperature and pressure.
The gravitational acceleration is $g = \SI{9.81}{m/s^2}$, 
yielding the dimensionless quantities $C = 0.0043$, and $\mathrm{Bd} = 76.456$, which quantify the ratios of viscous and surface tension forces to gravitational forces, respectively.
The properties of the lighter fluid are determined by the density and viscosity ratios, $\At$ and $\eta$.
\Cref{datasets} summarizes the configurations studied.
The equation of state parameters for the lighter fluid are $(\gamma, \pi_\infty) = (1.4,0)$ for the $\At=1$ case and $(\gamma, \pi_\infty) = (3.48, 1.46\times 10^7)$ for the $\At = 0.9$ case.
A startup period lasting a quarter of an oscillation cycle initiates fluid motion, ensuring consistency with the non-zero velocity fields predicted by linear theory.
Phase interfaces are identified by isolating the $\alpha = 0.5$ isosurface of the volume fraction field.

\begin{table}[ht!]
    \centering
    \caption{Configurations studied.}\label{datasets}
    {\small
    \begin{tabular}{l l r r c c c l}
    Cases & $h$[\unit{\milli\meter}] & $a^*$  &$g^*$	 & $\omega^*$  &($\At$, $\eta$) & $k^*$ &$\gamma^*_{U}$\,(S/H)   \\
    \midrule
    \Cref{fig:results_stableRT}~(b) & 26.87 & 10 & 1  &  24.6  & (1,0)  & 31.4 & 3.77\,(S) \\
    \Cref{fig:results_stableRT}~(c) & 21.81 & 15 & 1 & 29.6   & (1,0) &31.4   & 4.47\,(S)  \\
    \Cref{fig:results_stableRT}~(d) & 16.75 & 30  & 1 &  39.0  & (1,0) & 31.4 & 6.80\,(S) \\
    \Cref{fig:results_stableRT}~(e) & 12.95 & 30 & 1 & 45.7   & (1,0)  &31.4   &  3.61\,(S)\\
    \multirow{2}{*}{\Cref{fig:Growth_unstableRT}~(a)} & 13.00 & 16 & $-1$ & 33.9   & (0.9,0.1)  & 2.89  &  1.21\,(H) \\
       & 13.00 & 16 & $ -1$ & 33.9   & (0.9,0.1)  &29.1   &  1.21\,(S) \\
    \multirow{2}{*}{\Cref{fig:Growth_unstableRT}~(b)} & 13.00 & 18 & $-1$ & 27.4   & (0.9,0.1)  & 2.04  &  0.96\,(H) \\
           & 13.00 & 18 & $ -1$ & 27.4   & (0.9,0.1)  &24.5   &  3.45\,(S) 
    \end{tabular}
    }
\end{table}

\section{Results}\label{results}


% ...

\begin{figure*}[ht!]
    \begin{tikzpicture}
    \pgftext{\includegraphics{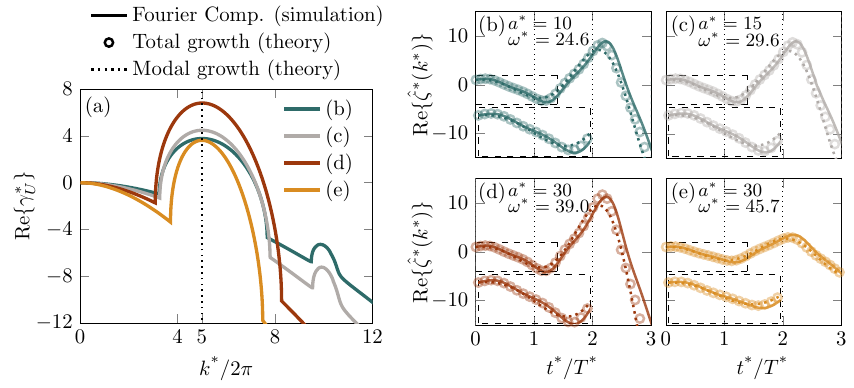}}
       \node at (3.7,3.4) {Parameters are provided in \cref{datasets}};
    \end{tikzpicture}
    \caption{
        Analysis of cases with the most unstable wavenumber $k^*/2\pi=5$ for $g^*=1$:
        (a) Growth rates;
        (b--e) Temporal evolution of the most unstable waves, with a magnified view showing the first period.
    }
    \label{fig:results_stableRT} 
\end{figure*}

We verify the linear theory predictions for the RT-stable case, where $g^*=1$.
In this scenario, the interface is modeled as an ideal free surface, with the lighter fluid assumed to have zero density and viscosity.
Consequently, only Faraday-type instabilities can be triggered via vertical vibration.
\Cref{fig:results_stableRT}~(a) shows the growth rates for four oscillation amplitude and frequency combinations, each with a positive maximum growth rate at wavenumber $k^*/2\pi = 5$.
This behavior indicates the onset of Faraday instability, leading to an unstable interface.
For all oscillation frequencies considered, the boundary layer thickness, $\delta \equiv \sqrt{2\nu_d / \omega}$, is sufficiently smaller than the fluid depth ($\delta / h \leq 0.019$)
and so the onset of the Faraday instability is subharmonic~\citep{kumar1996}.
The temporal evolution of the resulting Faraday waves over three oscillation periods is shown in \cref{fig:results_stableRT}~(b--e).
Initial transient dynamics of the Faraday waves are faithfully represented by the exponential modal growth of the most unstable Floquet exponent and its associated periodic behavior, with less than 1\% error compared to numerical simulations.
This agreement suggests that the nonmodal growth of Faraday waves is effectively represented.
These Faraday waves grow subharmonically and eventually lead to interface breakup.

\begin{figure*}[ht!]
    \includegraphics{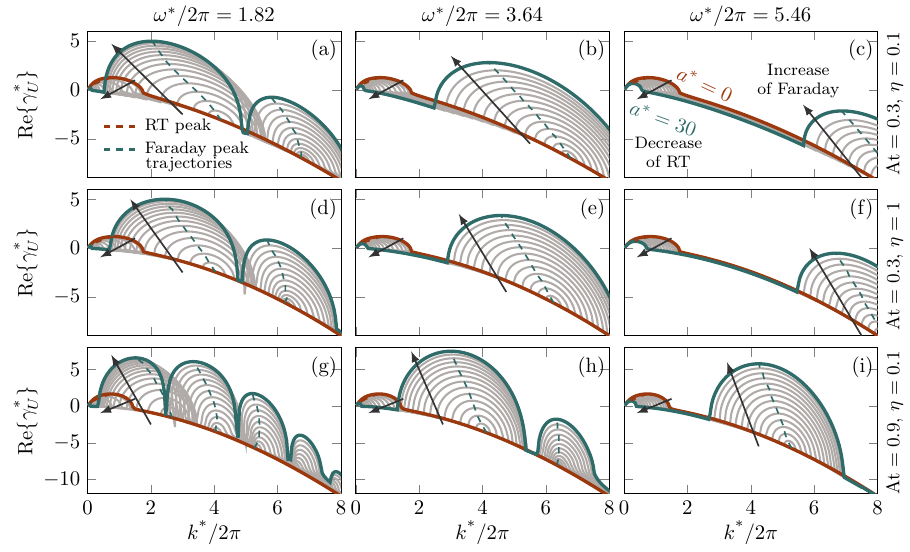}
    \caption{Interface displacement growth rate for $g^*=-1$ under various parameter combinations, with arrows denoting the transition from 
    $a^*=0$ to $a^*=30$.}\label{fig:Growthrate_unstableRT} 
\end{figure*}

We focus on the case where both the RT and Faraday instabilities coexist, specifically under a gravitational configuration directing gravity from the denser fluid toward the lighter one ($g^* = -1$).
\Cref{fig:Growthrate_unstableRT} shows the growth rates of the displacement of the interface for various combinations of parameters. 
The lighter fluid properties are chosen to approximate three types of interfaces: inviscid liquid--liquid ($\At=0.3$, $\eta=0.1$), liquid--liquid ($\At=0.3$, $\eta=1$), and gas--liquid ($\At=0.9$, $\eta=0.1$).
A classical RT instability manifests as a low-wavenumber peak in the zero-vibration case $a^*=0$.
For larger oscillation amplitudes, the amplitude of the positive peak diminishes toward zero and its position shifts toward the coordinate origin.
This suppression \textit{does not} lead to a complete stabilization of the RT instability in infinite domains, as the RT growth rate $\gamma_U^*$ remains positive, albeit reduced, under vibration.
In contrast, complete stabilization can be achieved in a finite horizontal domain, where the truncation of long-wavelength RT modes with $k^* \to 0$ effectively eliminates their growth.
This behavior demonstrates the theoretical principle of dynamic stabilization of RT waves under vibration within a confined container~\citep{wolf1969dynamic,wolf1970dynamic}.
Given its continuous evolution with increasing vibration, we refer to the instability mechanism of the low-wavenumber peak as RT-type.
The Floquet exponent associated with the RT peak is purely real and can thus be generalized to the cases of harmonic response.
The Faraday instability emerges as the vibration amplitude increases, marked by a distinct peak at higher wavenumbers that shifts downward.
The increasing prominence of the Faraday peak does not resonate with the RT peak.
Instead, the RT peak is suppressed and pushed towards the zero-wavenumber limit, further highlighting the competitive interplay between these two phenomena.
While an increase in the viscosity ratio, $\eta$, has little effect on the growth rate profiles, increasing the density difference, $\At$, induces local peaks at higher harmonics.
Unstable higher-order harmonics occur at the lowest oscillation frequency ($\omega^*/2\pi=1.82$) in \cref{fig:Growthrate_unstableRT}~(d, g), complementing the dominant Faraday subharmonic mode.
As expected, increasing the oscillation amplitude will eventually cause the Faraday peak to surpass the RT peak.
This transition between the two fundamental hydrodynamic instabilities leads to a mixed region where \textit{both instabilities coexist}, potentially altering the dynamics of the interface.

\begin{figure*}[ht!]
    \includegraphics{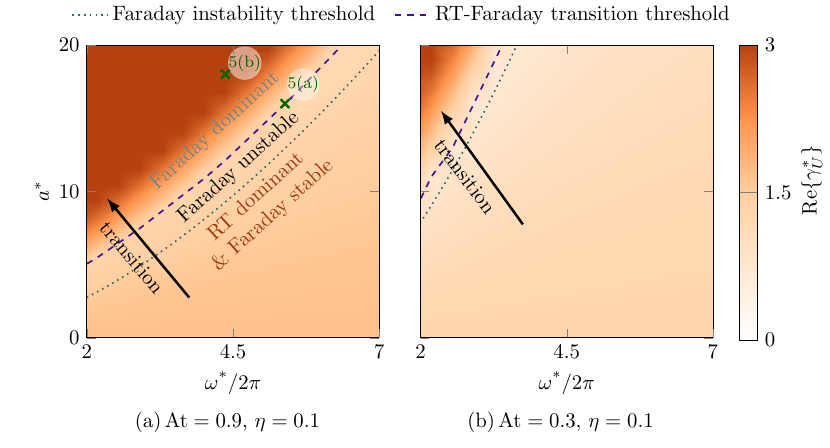}
    \caption{
        \label{fig:bifurcation_unstableRT} 
        Dominant growth rate across the oscillating frequency-amplitude phase space: 
        (a) $(\At,\eta)=(0.9,0.1)$; (b) $(\At,\eta)=(0.3,0.1)$.
        The margins for the Faraday instability and the RT-Faraday transition are highlighted.
    }
\end{figure*}

To have a comprehensive understanding of the multi-modal instability region and the associated transition mechanisms,
\cref{fig:bifurcation_unstableRT} shows the dominant growth rate within the oscillating frequency-amplitude phase space for two different density ratios.
For both cases, the transition of the leading instability mechanism from RT to Faraday can be observed in three stages.
At low oscillation amplitudes, the RT instability is the dominant phenomenon. 
The Faraday instability approaches its unstable boundary for larger oscillation amplitudes, yet the RT mechanism is dominant.
Finally, at sufficiently high oscillation amplitudes, the growth rate of the Faraday instability surpasses that of the RT instability.
For a given oscillation frequency, the thresholds of these two stages differ by about $2g^*$ for $\mathrm{At=0.9}$.
When the density difference between the two fluids is reduced ($\mathrm{At} = 0.3$), the RT instability dominates most of the phase space under consideration, albeit with a reduced growth rate.
The onset of Faraday instability requires higher oscillation amplitudes at the same oscillation frequency.
The threshold separating RT-dominated and Faraday-dominated regimes decreases to approximately $0.5g^*$.
These trends are consistent with what one expects; both instabilities weaken in the limit of vanishing density contrast.
In the following analysis, we focus on the case of $\mathrm{At} = 0.9$, where both instabilities are more strongly amplified compared to cases with lower density contrast.

\begin{figure*}[ht!]
    \includegraphics{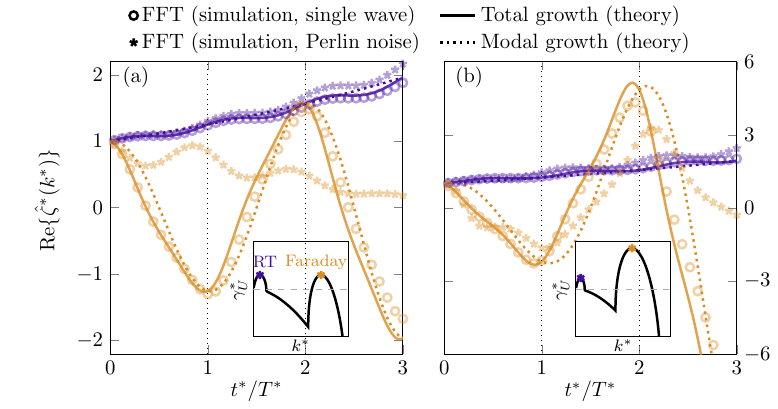}
    \caption{
        \label{fig:Growth_unstableRT} 
        Normalized evolution of Faraday and RT waves with growth rates (insets) under single-mode or Perlin noise excitations: (a) $a^*=16,\,\omega^*=33.9$; (b) $a^*=18,\,\omega^*=27.4$.
        Both are marked in \cref{fig:bifurcation_unstableRT}~(a).
    }
\end{figure*}

To quantitatively compare theoretical predictions and numerical simulations, we perturb the initial interface with three profiles: single-mode RT and Faraday sine waves and multiscale Perlin noise perturbations~\citep{Perlin2002}.
The temporal dynamics of the normalized RT and Faraday modes are shown in
\cref{fig:Growth_unstableRT} ~(a,\,b), corresponding to the two cases marked in \cref{fig:bifurcation_unstableRT}~(a).
While classical linear stability analysis of the static RT problem predicts purely exponential modal growth, vibrations induce an oscillatory pattern around this baseline growth.
Owing to its subharmonic nature, the Faraday displacement becomes negative at $t^*=T^*$.
The interface rebounds to a positive peak that either intersects the RT displacement (panel~a) or exceeds it (panel~b) at $t^*=2T^*$, where $T^*\equiv 2\pi/\omega^*$.
For perturbations with a single sine wave, the modal growth of the most unstable Floquet exponent, $\rme^{\gamma^*_U t^*}$, exhibits favorable agreement with the Fourier components obtained from simulations at integer multiples of the oscillation period for both instabilities.
This alignment confirms that the interface growth follows Floquet theory.
By incorporating the modal growth with the additional periodic terms,
the total growth defined in Eq.~\cref{eqn:Floequet_form} accurately predicts the interfacial dynamics during the first oscillation period.
Beyond this period, the theory remains in qualitative agreement with simulations, although minor deviations arise due to nonlinear effects.
Simulations initialized with Perlin noise show suppressed growth rates and reduced amplitudes of Faraday waves compared to theoretical predictions. 
The broadband initial wavenumber spectrum induces a harmonic-like response in the Faraday waves when their growth rate matches that of the RT instability.
Without competing wavelengths, these waves would exhibit subharmonic Floquet-type growth and ultimately induce interface breakup, suggesting that the observed suppression is primarily attributable to nonlinear damping effects.
When the Faraday growth rate exceeds that of RT, the theory remains qualitatively predictive of the system’s dynamics, despite multiscale waves.
In contrast, the growth of RT waves is generally unaffected by the presence of multiscale wave components.


\begin{figure*}[ht!]
    \includegraphics{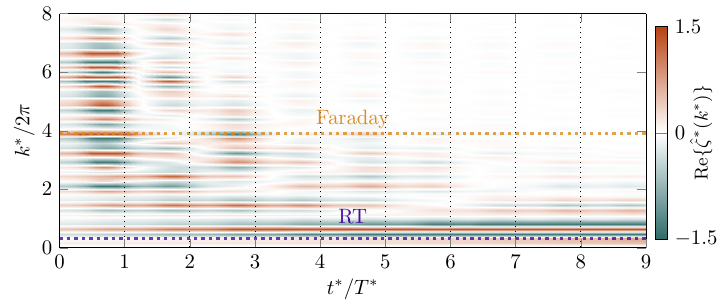}
    \caption{
        \label{fig:Wave_perlin} 
        The wavenumber–time diagram for the Perlin noise case in \cref{fig:Growth_unstableRT}~(b).
    }
\end{figure*}

\begin{figure*}[ht]
    \includegraphics{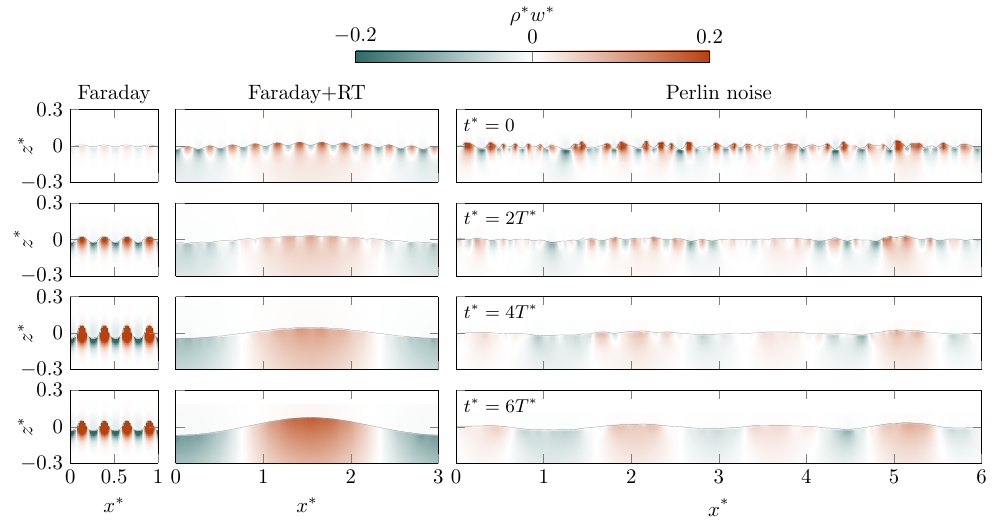}
    \caption{
        Temporal evolution of the vertical momentum field and interface profile.
    }
    \label{fig:damped Faraday} 
\end{figure*}

To illustrate transient interface dynamics across wavenumbers in a realistic context, \cref{fig:Wave_perlin} presents the wavenumber–time diagram for the Perlin noise case shown in \cref{fig:Growth_unstableRT}~(b).
This case exemplifies the nonlinear damping of Faraday waves despite a dominant linear growth rate.
The coexistence of RT and Faraday waves is observed within the given time window.
Wavenumbers near the Faraday peak exhibit periodic phase changes due to their dominance of the subharmonic nature, whereas those near the RT peak display steady growth.
Initially ($t\lesssim2.5T^*$), the interface profile is dominated by Faraday waves due to their higher initial growth rate.
Long-wavelength RT waves amplify and dominate as time progresses, though residual small-scale Faraday oscillations persist. 
This transition implies that the steadily growing RT components nonlinearly dampen the oscillatory subharmonic Faraday responses, despite the inherent instability of these latter responses.
The resulting interface profiles, shown later in \cref{fig:damped Faraday}, provide a direct visualization of the competition between long-wavelength RT waves and short-wavelength Faraday waves.
\Cref{fig:Growth_unstableRT}~(a,\,b) shows that the Faraday waves would otherwise grow unimpeded toward interface breakup without RT competition.
The dominance of RT waves suppresses this pathway, redirecting the system toward RT-driven dynamics.
Together, these results demonstrate the nonlinear transient dynamics that govern fluid interfaces under the combined effects of density and vibration.

To better understand the nonlinear damping of Faraday waves that cannot be represented via linear theory, we examine the evolution of the entire flow field, rather than the interface alone.
\Cref{fig:damped Faraday} shows the vertical momentum field, $\rho^*w^*$, a conservative quantity, at various time instants.
A reference case is also considered for direct comparison, in which a linear superposition of Faraday and RT modes with equal amplitude perturbs the initial interface.
The velocity fields exhibit wavenumber patterns consistent with those observed at the interface.
Faraday-type wave packets remain localized near the interface, while RT-type wave packets penetrate deeper, occupying the entire lower half of the domain.
These spatial structures reflect their underlying physical mechanisms: the former arises from interfacial resonance, while buoyancy-induced pressure gradients drive the latter.
Faraday waves are readily developed in the absence of competing length scales.
Their nonlinear damping does not require broadband perturbations such as Perlin noise.
Including a single additional RT mode is sufficient to suppress the Faraday response.
As discussed in the context of \cref{fig:Growthrate_unstableRT},
increasing the vibration amplitude to amplify the Faraday instability will consistently suppress the RT instability toward the zero-wavenumber limit and reduce its growth rate.
Numerical simulations further reveal that the residual, weakened RT dynamics concurrently attenuate the Faraday response through a nonlinear damping mechanism.
These results demonstrate a bidirectional competitive interaction between the two fundamental instabilities, each modulating the other through linear and nonlinear processes.








\section{Conclusions}\label{conclusion}

This manuscript presents a theoretical and computational investigation into the previously unexplored coexistence and competition of two fundamental hydrodynamic instabilities: the pressure gradient-driven Rayleigh--Taylor (RT) instability and the resonance-induced Faraday instability.
Through linear Floquet stability analysis and 2D numerical simulations, we demonstrate that interface growth under combined density and vibration effects exhibits a Floquet-type behavior: total displacement decomposes into modal growth of the most unstable Floquet mode and harmonic oscillations from periodic terms.
In an unbounded horizontal domain, we show that vibration can mitigate the growth rate of RT instabilities.
This result aligns with the dynamical stabilization of RT instabilities in confined containers, where physical boundaries restrict long-wavelength modes.
We also reveal a previously unseen multi-modal instability region, where the RT and Faraday instabilities coexist and compete in non-trivial and unintuitive ways.
This region emerges as the oscillation amplitude increases, marking a gradual transition in the dominant instability mechanism from RT to Faraday.
Within this region, we observed a unique bidirectional competitive interaction.
The growing Faraday mechanism, amplified by increasing vibration amplitude, consistently suppresses the RT instability toward the zero-wavenumber limit.
In turn, the residual RT mechanism attenuates the Faraday responses via a nonlinear damping mechanism, even when the initial growth rate of the Faraday instability exceeds that of RT.
Furthermore, the coexistence of these mechanisms generates multiscale interfacial waves, culminating in interfacial breakup phenomena that are ubiquitous in both natural and engineered systems, yet remain uncharted in prior studies.

Although linear stability analysis predicts a multi-modal instability region where RT and Faraday instabilities should coexist, numerical simulations reveal a critical nuance.
While each instability develops independently at its theoretically predicted growth rate, sustained coexistence is difficult to achieve due to the nonlinear damping of Faraday waves by RT-induced flows.
This bidirectional competition offers a plausible explanation for the absence of simultaneous RT and Faraday instabilities in most natural settings.
Nonlinear approaches, such as weakly nonlinear stability analysis, offer a promising way to account for these cross-mode interactions.
Extending the analysis and simulations to 3D, although more computationally demanding, enables the capture of more complex interfacial instability patterns, such as squares and hexagons.
These structures provide a more faithful representation of real-world scenarios.
Nevertheless, the current 2D approach effectively captures the competition between the instability mechanisms at different scales and remains valid under geometric symmetries that permit dimensionality reduction, such as azimuthal decomposition in cylindrical containers.

The discovery of coexistence and competition between RT and Faraday instabilities offers new insights into predicting and controlling fluid interfaces in vibration-prone systems.
Our findings bridge fundamental hydrodynamics and engineering, leveraging instability interactions of the fluid to control the interface kinematics.
Vibrations dynamically stabilize RT modes while exciting Faraday instabilities, and the inclusion of RT waves, in turn, nonlinearly dampens the unstable Faraday response.
By tuning the oscillation, processes like atomization or inertial confinement fusion can be optimized to suppress both low- and high-wavenumber instabilities, thereby enhancing efficiency and preventing failure.
These insights reveal new avenues for studying transient dynamics in natural and engineering flow systems, where external vibrations compete with intrinsic instabilities, ranging from geophysical flows to the injection of bubbles.

\section*{Acknowledgments}

Sandia National Laboratories is a multi-mission laboratory managed and operated by National Technology \& Engineering Solutions of Sandia, LLC (NTESS), a wholly owned subsidiary of Honeywell International Inc., for the U.S. Department of Energy’s National Nuclear Security Administration (DOE/NNSA) under contract DE-NA0003525. 
This written work is authored by an employee of NTESS. The employee, not NTESS, owns the right, title and interest in
and to the written work and is responsible for its contents. Any subjective views or opinions that might be expressed in the written work do not necessarily represent the views of the U.S.\ Government.
The publisher acknowledges that the U.S.\ Government retains a nonexclusive, paid-up, irrevocable, world-wide license to publish or reproduce the published form of this written work or allow others to do so, for U.S.\ Government purposes. The DOE will provide public access to results of federally sponsored research in accordance with the \href{https://www.energy.gov/downloads/doe-public-access-plan}{DOE Public Access Plan}.
This work used the resources of the Oak Ridge Leadership Computing Facility at the Oak Ridge National Laboratory, which is supported by the Office of Science of the U.S. Department of Energy under Contract No. DE-AC05-00OR22725 (PI Bryngelson, allocation CFD154).
This work also used Bridges2 at the Pittsburgh Supercomputing Center through allocation TG-PHY210084 (PI Spencer Bryngelson) from the Advanced Cyberinfrastructure Coordination Ecosystem: Services \& Support (ACCESS) program, which is supported by National Science Foundation grants \#2138259, \#2138286, \#2138307, \#2137603, and \#2138296.
The authors thank F.\ Pierce and J.\ S.\ Horner for helpful discussions about this work.


\section*{Data availability}

Code available at \url{https://github.com/MFlowCode/MFC}.




\appendix
\setcounter{equation}{0}
\renewcommand{\theequation}{{\rm A}.\arabic{equation}}

 \setcounter{table}{0}
\renewcommand{\thetable}{{\rm A}.\arabic{table}}

  
\section*{Appendix}


Substituting the general solution in Eq.~\cref{eqn:w_genearal}
into the boundary conditions in Eqs.~\Crefrange{eqn:BC_vertical}{eqn:BC8}
yields a linear system comprising eight boundary conditions:
\begin{align}
    &a_{n,l}+ b_{n,l}+ c_{n,l}+ d_{n,l}= a_{n,d}+ b_{n,d}+ c_{n,d}+ d_{n,d}=\gamma_n^*\hat{\zeta}_n^*.\\
   &a_{n,l}- b_{n,l}+ q_{n,l}c_{n,l}- q_{n,l}d_{n,l}= a_{n,d}- b_{n,d}+ q_{n,d}c_{n,d}- q_{n,d}d_{n,d}.\\
  & \eta\left(2a_{n,l}+ 2b_{n,l}+ (q_{n,l}^2+1)c_{n,l}+ (q_{n,l}^2+1)d_{n,l}\right)\nonumber\\
  &= 2a_{n,d}+ 2b_{n,d}+ (q_{n,d}^2+1)c_{n,d}+ (q_{n,d}^2+1)d_{n,d}.\\
     &a_{n,l}\rme^{k^*}+ b_{n,l}\rme^{-k^*}+ c_{n,l}\rme^{ k^*q_{n,j}}+ d_{n,l}\rme^{-k^*q_{n,j}}=0\\
    & a_{n,d}\rme^{-k^*}+ b_{n,d}\rme^{k^*}+ c_{n,d}\rme^{-k^*q_{n,j}}+ d_{n,d}\rme^{k^*q_{n,j}}=0\\
    & a_{n,l}\rme^{k^*}- b_{n,l}\rme^{-k^*}+ q_{n,l}c_{n,l}\rme^{ k^*q_{n,l}}-q_{n,l}d_{n,l}\rme^{-k^*q_{n,l}}=0\\
   &  a_{n,d}\rme^{-k^*}- b_{n,d}\rme^{k^*}+ q_{n,d}c_{n,d}\rme^{-k^*q_{n,d}}-q_{n,d}d_{n,d}\rme^{k^*q_{n,d}}=0
\end{align}
Upon simplification, this system can be compactly represented in matrix form using the matrix $\vb*{Q}_n$, defined as
\begin{align}\label{eqn:Q}
    \vb*{Q}^*_n \equiv 
    \mqty(
   1 & 1 & 1 & 1 & 0 & 0 &0 &0 \\
   1 & 1 & 1 & 1 & -1 & -1 & -1 &-1 \\
   1 & -1 & q_{n,d} & -q_{n,d} & -1 & 1 & -q_{n,l} &q_{n,l} \\
   2 & 2 & 1+q_{n,d}^2 & 1+q_{n,d}^2 &-2\eta &-2\eta & -\eta (1+q_{n,l}^2)  & -\eta (1+q_{n,l}^2) \\
   \rme^{-k^*q_{n,d}^+} & \rme^{-k^*q_{n,d}^-} & \rme^{-2k^*q_{n,d}^+}& 1 & 0&0 &0 & 0 \\
   q_{n,d}^+\rme^{-2k^*} & -q_{n,d}^- &2q_{n,d} \rme^{-k^*q_{n,d}^+}& 0 & 0&0 &0 & 0 \\
     0 & 0& 0 &0 & \rme^{-k^* q_{n,l}^-} & \rme^{-k^* q_{n,l}^+} & 1 & \rme^{-2k^*q_{n,l}}   \\
     0 & 0& 0 &0 & -q_{n,l}^-  & q_{n,l}^+ \rme^{-2k^*} & 0 & 2\rme^{-k^*q_{n,l}^+} 
    ).
\end{align}
Here, the superscripts $\left(\cdot\right)^+$ and $\left(\cdot\right)^-$ are used to denote $q^{+} \equiv q+1$ and $q^{-} \equiv q-1$.
The boundary conditions are thus compactly written as
\begin{align}
     &{\vb*{Q}_n}
\mqty(a_{n,d}, b_{n,d},c_{n,d},d_{n,d},a_{n,l}, b_{n,l},c_{n,l},d_{n,l})^\transpose  = 
     \mqty(\gamma^*_n \hat{\zeta}^*_n, 0,0,0,0,0,0,0)^\transpose.
\end{align}
Eq.~\Cref{eqn:Q} can be directly employed to solve the linear system in various limiting cases, such as small or large wavelengthsextreme viscosity and density ratios, and large domain heights, without introducing spurious numerical instabilities.
This simplification thus enables a comprehensive analysis of interfacial dynamics across varying density contrasts, viscous effects, and domain sizes.



\bibliography{ref.bib}



\end{document}